# Estimator Model For Prediction Of Power Output Of Wave Farms Using Machine Learning Methods

Bhavana Burramukku
Master of Data science
The University of Adelaide
Adelaide, Australia
a1789278@student.adeliade.edu.au

## 1. Abstract

The amount of power generated by a wave farm depends on the Wave Energy Converter (WEC) arrangement along with the usual wave conditions. Therefore, forming the appropriate arrangement of WECs in an array is an important factor in maximizing power absorption. Data collected from the test sites is used to design a neural model for predicting wave farm's power output generated. This paper focuses on developing a neural model for the prediction of wave energy based on the data set derived from the four real wave scenarios from the southern coast of Australia. The applied converter model is a fully submerged three-tether converter called CETO. A precise analysis of the WEC placement is investigated to reveal the amount of power generated by the wave farms on the test site.

## 2. Introduction

Today, one of the world's most pressing challenges is the generation of energy without adverse environmental impacts. Traditional ways of generating energy, such as fossil fuels, are not sustainable and would reduce their usage over time. Therefore, for this reason, renewable energy resources are the best alternatives. In certain countries, renewable energy resources are widely used, such as solar, wind, wave energy. As renewable energy resources are more expensive than traditional sources of energy, such as fossil-fuel sources, in the past, governments have not been interested in these energy resources. Fortunately, this perspective has shifted lately. For example, by 2004, despite the growing usage of wind power, the cost of converting energy from wind power had decreased by about 80 percent over the previous 30 years as per the National Energy Policy Commission, 2004. According to [1] wave energy is a resource of renewable energy with immense power capacity and minimal environmental consequences. Wave energy extraction depends not only on the form of wave energy converters but also on the location of the wave farm.

Research shows that a physics-based model or a statistical model can be used to develop a prediction model. Physics-based models have existed since the 1960s, with the hypothesis emerging in the late 1950s.[2][3][4] In the 1990s, time series models began, and the neural networks became the most commonly used approach. Recent research proposed a model incorporating both of these two methods by statistical postprocessing of physics-based model forecasts [2]. Wave forecasting models have become more robust and highly accurate after decades of study. These models, however, need massive quantities of data to train, which, even for a small-scale prediction, consumes a lot of time. For this purpose, algorithms based on machine learning and deep learning have been designed and developed to increase the accuracy and speed of prediction.

One of the pre-requisites for any prediction is the creation, calibration, and validation of an aptly efficient approach. In addition, this model needs to be able to operate reasonably quickly and incorporate suitable forecast data into its forecasts. This study uses a neural network-based machine learning model to predict the accuracy of the power output based on WEC 's positions and absorbed power output from the four real wave farms on Australia's southern coast.

## 3. Literature Review

Wave forecasting techniques are usually divided into two categories: process-based models and statistical techniques. Process-based models have been developing since the 1950s and are based on theoretical and experimental research [2][3][4]. Recent developments in this field include the advent of models of the third generation (3 G), such as WAM (Wave Modelling) used for deep-water wave modeling [5] and SWAN (Simulating Waves Nearshore) for high-resolution grids to achieve cost-effective modeling outcomes in coastal areas[6-7]. Statistical techniques are a new improvement in the field. Artificial neural networks (ANNs) are one of the most practiced approaches [8-12].

ANN has been extensively applied in ocean engineering [13] for predicting tsunami water levels. To train the ANN, they used various tsunami conditions and then predicted the preliminary and secondary tsunami waves that corresponded well with the numerical tsunami simulation results. There are also several studies done in which ANN has been used to estimate the parameters of waves. [14] demonstrated that ANN successfully predicts wave parameters in a short runtime, and the predicted values are very similar to the values observed. To estimate the wave heights and times,[15] used the neural network method and found that the predicted wave parameters fit well with those hindcasted using Young's model. In order to forecast significant wave heights and zero-up-crossing wave intervals, [16] used two separate neural network techniques and demonstrated the suitability of the artificial neural technique for short-term wave parameter forecasts.[8] used multiple linear regression (MLR) and ANN models for significant wave height prediction and found that ANN provides smaller MSEs for this parameter prediction.

Wave forecast models can be of high computational cost. A novel approach with machine learning [17] is discussed here. The objective of this approach is to train machine learning models on several outcomes of a physics-based wave model forced to accurately reflect wave conditions. Accurate power output forecasts are enabled by forecasting these wave conditions at locations corresponding to a (potential) wave energy converter (WEC) array. Given the recent evolution of a distribution scheme for wave energy resource, the power potential for a hypothetical WEC array can be calculated if the wave conditions at a specific position can be predicted.[18]

The cost of computing is often a significant limitation of real-time forecasting systems [19]. Here, we use machine learning techniques to predict wave conditions to substitute a computationally intensive physics-based model as machine-learning techniques can predict wave conditions for a fraction of the computational cost with similar accuracy to a physics-based model. Machine learning has been used to predict wave conditions in various studies [12][20]. Immense appetite for data is one of the obstacles for machine learning applications. It is more of an anomaly than a rule that a machine learning approach has to have an optimal amount of data. However, there is the luxury of being able to run the physics-based model as many times as possible to create a sufficiently large data set to train the machine learning model while developing a machine learning surrogate for a physics-based model. We describe a surrogate model here [21] as a data-driven technique to empirically approximate a physics-based model's response surface. Alternatively, these were called "metamodels", "model emulators", and "proxy models" [18]. The inputs and outputs of several thousands of wave model outputs were accumulated into a suitably large collection of input data to assemble the training dataset needed to create a robust machine learning model.

Depending on the configuration of the array, the spatial dispersion of the WEC in the array and the interactions between them can have a positive (wave intensity, smoothing of the output power [22],[23]) or a negative effect (masking effects). It is therefore of great importance to be able to anticipate these interactions in order to use or prevent them in a beneficial way. In the past, this has been the subject of research by many R&D units across Europe, see the pioneering work of [24], [25], and it is still a research field, as several recently published studies indicate, see [26],[27].

## 4. Data Collection And Preprocessing

The applied wave energy dataset in this research is collected and published by [28][33]. In order to develop an estimator model, the applied dataset from [29] has been considered. Wave Energy Converters (WEC) data set consists of positions and the absorbed power outputs of WECs in four real wave scenarios from the southern coast of Australia (Sydney, Adelaide, Perth, and Tasmania). Recently, a considerable number of optimisation investigations is performed on these locations in order to propose the best WECs arrangement in a wave farm [34-38]. The applied converter model is a fully submerged three-tether converter called CETO [30]. 16 WECs locations are placed and optimized in a size-constrained environment. The data set consists of 48 attributes, out of which the first 32 attributes are the positions (latitude and longitude) of the 16 WEC, continuous from 0 to 566 (m). The next 16 attributes are WECs absorbed power. The last attribute is the total power outputs from the farms. The WECs positions are used to predict the total power outputs. A total of 72000 rows of data is present in each dataset. Each row consists of the data from the wave farm by changing the positions of the 16 converters and their respective absorbed power is noted. For validation purposes 20% of the data from each dataset is considered as the test data and the remaining 80% data is used for training the model. The neural model developed is trained individually for each dataset.

The gap distance between the converters and the angle between converters is known to influence the power output. As we have the WEC position information, we have calculated average distance and mean distance of the converters in order to analyze the influence of distance on power generation. Hence, we calculated the average distance of a single converter from the remaining 15 converters. Likewise, average distance is calculated for all the 16 WEC in the wave farm. Once the average distance between converters is calculated, the mean of all the average distances is calculated in order to analyze the relationship between distance and power. Below figure 1 shows the plot for the mean distance and total absorbed power for Adelaide wave farm.

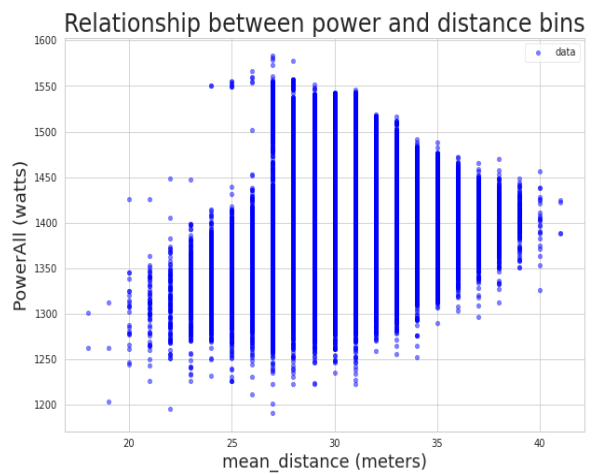

Figure 1 Relationship between distance and total absorbed power

Statistical Analysis is the process of collecting, exploring, and organizing data, exploring patterns and trends using various descriptive, Predictive, Exploratory data analysis techniques. In general, there are three fundamental tasks of statistical learning from data: classification, regression, and probability density modeling. As usually, exploratory data analysis (EDA) is the first step of the study along with the visualization of data. Figure 3 shows the distribution of distance and total absorbed power. There are 3 major tasks

1. Feature extraction and feature engineering: Transformation of raw data to make them suitable for modeling

2. Feature Transformation: transformation of data to improve the accuracy of algorithm

3. Feature selection: removing unnecessary data. Data handling like removing the outliers if any and replacing missing values helps in noise reduction and better accuracy of the model.

As a part of pre-processing scaling and normalization of data is done before passing the data to the model. No outliers were removed from the dataset. Although there have been instances in Tasmania and Perth data where minimum distances are producing higher power values. But they seem to be like Natural Variation Outliers rather than data processing or sampling errors. Applied LOF, Z-score and box plots to identify possible outliers but the aforementioned extreme values were not identified by the algorithms as outliers. Extreme values occurred have lower probabilities and unusual, they are a normal part of the data distribution. Figure2 shows the LOF outlier algorithm

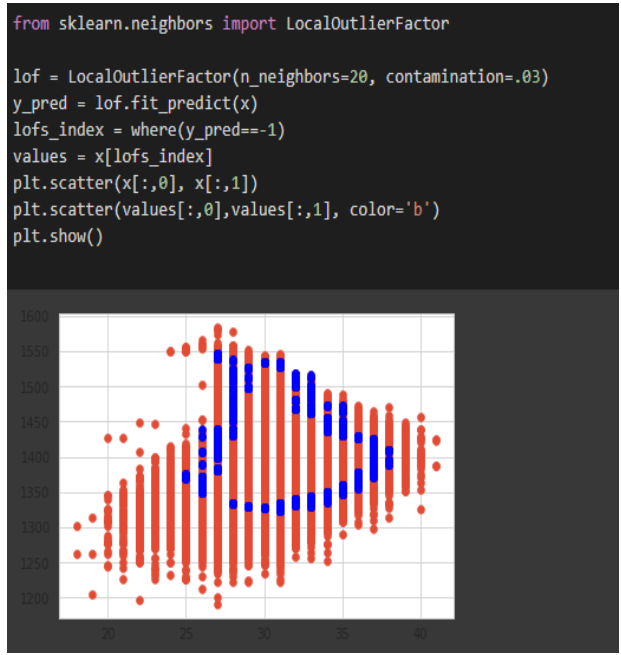

Figure 2 Data visualization plot for clusters after PCA

Local Outlier Factor (LOF) the algorithm tries to find anomalous data points by measuring the local deviation of a given data point with respect to its neighbours. In this algorithm, LOF would yield a score that tells if our data is an outlier or not.

LOF(k) ~ 1 means Similar density as neighbours.

LOF(k) < 1 means Higher density than neighbours (Inlier/not an outlier).

LOF(k) > 1 means Lower density than neighbours (Outlier)

## 5. Methodology : Neural Model

The primary purpose of the project is to develop an estimator model for the prediction of power output of wave farms using a machine learning framework based on recurrent deep neural networks. In order to compare the outcome of the neural model a base line neural model is developed using Multi-Layer Perceptron (MLP) framework.

MLP is one of the most common neural network models used in the field of machine learning. The basic structure of MLP consists of interconnected neurons transferring information to each other, much like the human brain. Each neuron is assigned a value. The network mainly consists of 3 layers the input layer, hidden layer(s), and output layer. The MLP is a feedforward neural network, which means that the data is transmitted from the input layer to the output layer in the forward direction. Initially, a framework with minimal layer of MLP will be developed and the accuracy of the power model is determined.

MLP model is developed for each of the dataset individually. Data is partitioned 80-20 for training and validation of model. Scaled and normalized data is passed to the MLP regressor. Below figure [4,5] shows the Keras MLP regressor model. Figures [6,7] shows the MSE plot for the Adelaide and Perth dataset. MLP model generated high accuracy when trained individually for the four real wave scenarios but gave very less accuracy when trained data combined from the four wave farms.

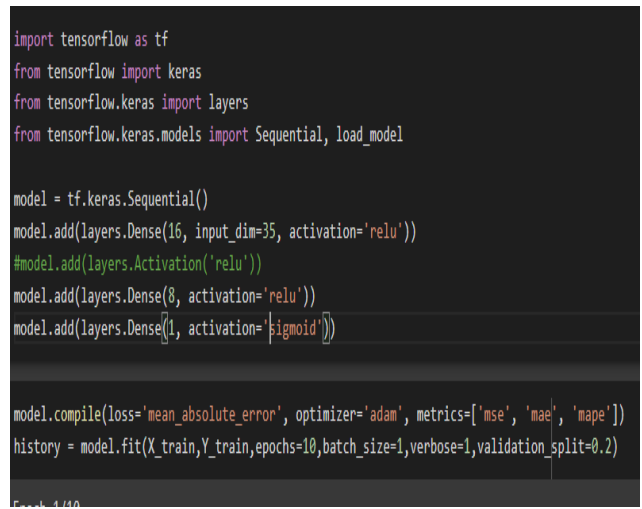

Figure 4 MLP Model

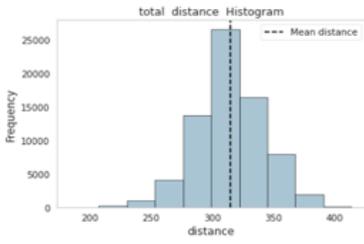 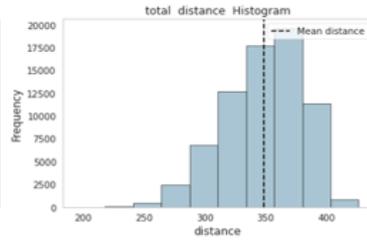 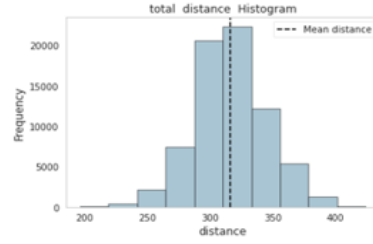 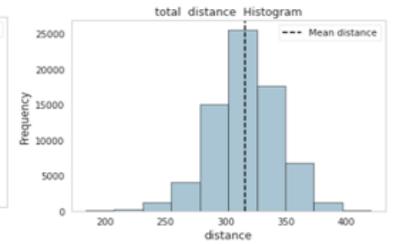
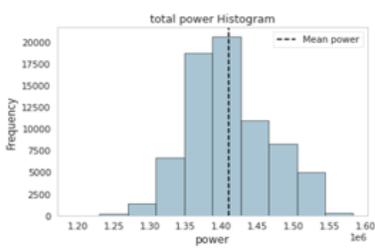 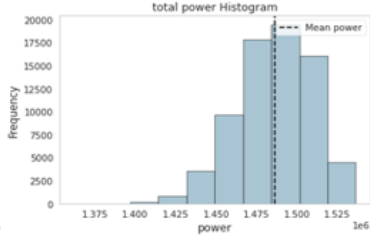 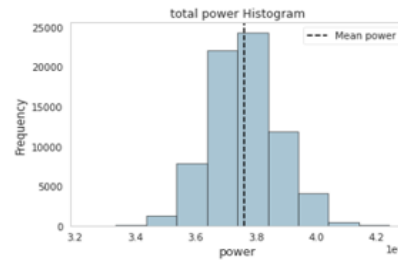 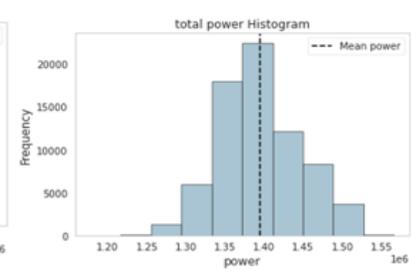

Figure3 Distance and power distribution in wave farms

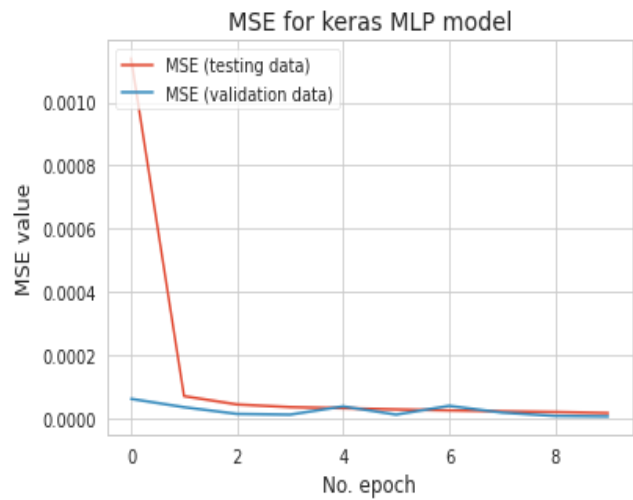

Figure 6 MSE metric for Adelaide data

Figure 5 Mean absolute error

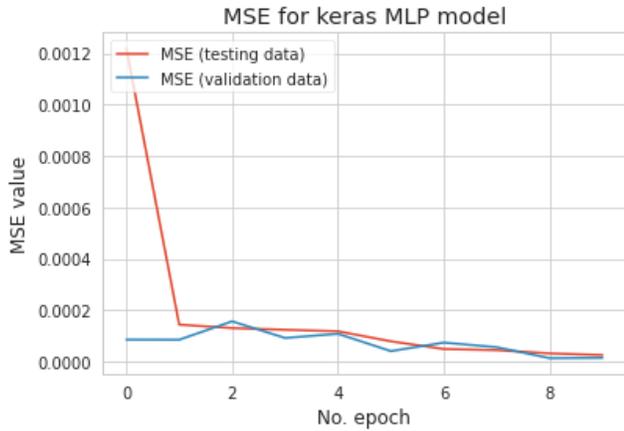

Figure 7 MSE metric for Perth data

Below table shows the comparison between the four wave scenarios.

| Wave farm | Data model | Correlation | Metrics |
|---|---|---|---|
| **Sydney** | Linear Data | Very Strong Correlation | Mean Distance: - 348.095701<br>Mean Power: - 1486229<br>Min Distance: - 194.7864<br>Max Distance: - 426.243416<br>Min Power: - 1361962<br>Max Power: - 1536347 |
| **Adelaide** | Non-linear Data | Moderate Correlation | Mean Distance: - 314.812861<br>Mean Power: - 1410073<br>Min Distance: - 184.652449<br>Max Distance: - 413.760604<br>Min Power: - 1191378<br>Max Power: - 1583052 |
| **Tasmania** | Non-linear Data | Not much Correlation | Mean Distance: - 316.294839<br>Mean Power: - 3760137<br>Min Distance: - 196.247859<br>Max Distance: - 424.120062<br>Min Power: - 3235131<br>Max Power: - 4241838 |
| **Perth** | Non-linear Data | Not much Correlation | Mean Distance: - 316.176215<br>Mean Power: - 1394474<br>Min Distance: - 183.90566<br>Max Distance: - 420.3663<br>Min Power: - 1177711<br>Max Power: - 1565836 |

Table 1 Comparison Chart

## 6. Findings

Wave forecasting is a non-linear problem, so linear regression is not suitable for this prediction model. The stochastic and non-linear interactions associated with the generation and integration of wave energy complicate effective modeling. Therefore, recurrent deep neural networks are the best suited model for this research.

Below are some of the finding from the statistical analysis of the data.

- In Sydney wave farm, as the distance increases the power generation also increases. Total power and mean distance are positively correlated.

- For the Adelaide and Perth farms, power values increase approximately till the mean average distance and later as the distance increase, power value are reduced.

- Tasmania wave farms, power values increase approximately till the mean average distance and later as the distance increase, power value are reduced. But as the average distance reaches 350m there seems to be a consistent power value.

- Tasmania wave farm has a higher power generation compared to the other locations which is almost double the power generated from Sydney, Adelaide, and Perth.

- No outliers were removed from the dataset.

- All though there have been instances in Tasmania and Perth where minimum distances are producing higher power values. But they seem to be Natural Variation Outliers rather than data processing or sampling errors.

- Applied LOF, Z-score and box plots to identify possible outliers but the aforementioned extreme values were not identified in the algorithms.

- Extreme values occurred have lower probabilities. They are a normal part of the data distribution.

- WEC in Perth wave farm are closed placed(clusters) whereas in all other location converters are spread throughout the farm. Below scatter plot shows WEC positions for 16 WEC over 10 combinations in Perth farm.

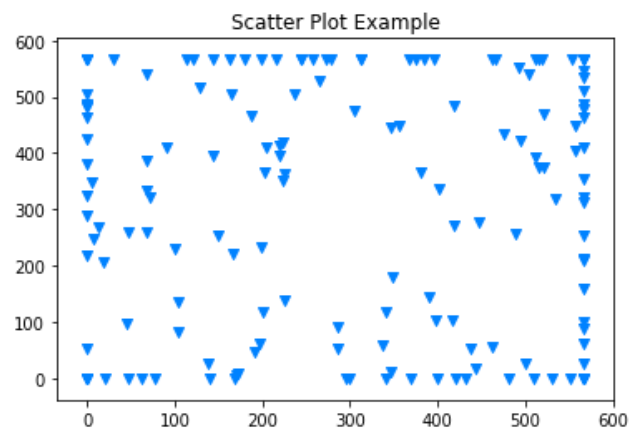

Figure 8 Sydney wave farm

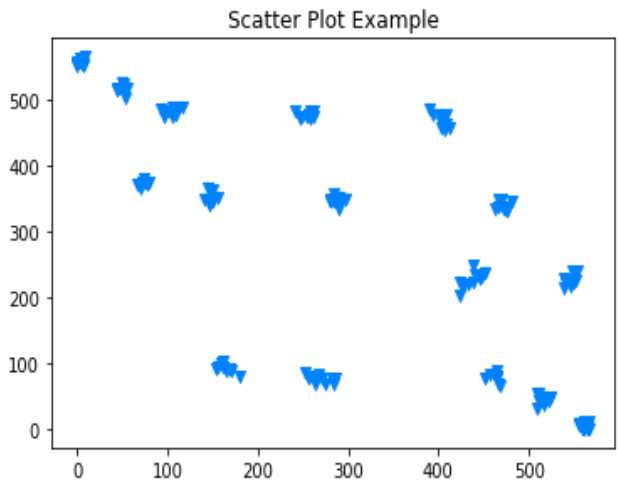

Figure 9 Perth wave farm

## 7. Future Directions

This MLP model is a basic neural model. Later an estimator model on recurrent deep neural networks will be developed. Recurrent deep neural networks (RNN) is one of the most common neural network models used in the field of deep learning and is a generalization of feed forward networks that has internal memory. RNNs are recurrent because they perform the same task for every input of a sequence, with the output being dependent on the earlier computations. RNNs have an internal memory that captures information about what has been processed so far. In the theoretical sense, RNNs can make use of data in arbitrarily long sequences, but in practice, they can look back to only a few steps. [31]

Long Short-Term model Neural Networks (LSTM) is a type of RNNs competent in learning long-term dependencies more accurately than conventional RNNs. The model consists of an input layer, hidden layers, and an output layer with each hidden layer of traditional RNN containing one short-term memory vector. LSTM retains information for long periods due to their inner cells which can carry data unchanged. LSTM has complete control over the cell state and can add, edit, or remove information in the cell using specialized structures called gates.

Recurrent networks rely on an extension of backpropagation called Back Propagation through Time (BPTT) to accurately classify sequential input. The use of backpropagation recurrent networks improves the prediction results. A recent study using the Adaptive Neuro-Surrogate Optimization (ANSO) model revealed that training and tuning the model with the LSTM network yielded better optimization results when compared with other methods [32].

Further research is summarized as below:

- Currently a simple Multi –Layer Perceptron model is developed to predict the output. The neural model is developed for each dataset and trained separately as each wave farm is having different characteristics.
- Next approach is Transfer Learning where pre-trained model can accommodate all kinds of wave scenarios. (Storing knowledge gained while solving one problem and applying it to a different but related problem.)
- Developing Recurrent Neural Networks (RNN) models like **Long Short-Term Memory Networks (LSTM)** etc. in order predict total power generated from the wave farms with less computational cost and more accuracy.
- Test and compare the performance of different deep learning models.